\documentclass[12pt]{article}
\input epsf.sty
\usepackage{mathrsfs}
\usepackage{amsmath}
\usepackage{mathrsfs}
\usepackage{amssymb}
\usepackage{color}
\usepackage{psfrag}
\usepackage{graphicx}
\usepackage{subfigure}
\usepackage{rotating}

\newcommand{\bea}{\begin{eqnarray}}
\newcommand{\eea}{\end{eqnarray}}
\newcommand{\be}{\begin{equation}}
\newcommand{\ee}{\end{equation}}
\newcommand{\vs}[1]{\vspace{#1 mm}}

\newcommand{\dsl}{\pa \kern-0.5em /}

\newcommand{\pa}{\partial}

\newcommand{\nn}{\nonumber\\}

\begin{document}
\topmargin 0pt
\oddsidemargin 0mm

\begin{flushright}



\end{flushright}

\vspace{2mm}

\begin{center}
{\Large \bf Calculating the jet quenching parameter in the plasma of 
NCYM theory from gauge/gravity duality}

\vs{10}

{Somdeb Chakraborty\footnote{E-mail: somdeb.chakraborty@saha.ac.in} and 
Shibaji Roy\footnote{E-mail: shibaji.roy@saha.ac.in}}

 \vspace{4mm}

{\em

 Saha Institute of Nuclear Physics,
 1/AF Bidhannagar, Calcutta-700 064, India\\}

\end{center}

\vs{10}

\begin{abstract}
A particular decoupling limit of non-extremal (D1, D3) brane bound state 
system of type IIB string theory is known to give the gravity dual of 
space-space non-commutative Yang-Mills (NCYM) theory at finite temperature. 
We use a string probe in this background to compute the jet quenching 
parameter in a strongly coupled plasma of hot non-commutative Yang-Mills 
theory in (3+1)-dimensions from gauge/gravity duality. We give expressions 
for the jet quenching parameter for both small and large non-commutativity. 
For small non-commutativity, we find that the value of the jet quenching 
parameter gets reduced from its commutative value. The reduction is 
enhanced with temperature as $T^7$ for fixed non-commutativity and  
fixed 't Hooft coupling. We also give an estimate of the correction due 
to non-commutativity at the present collider energies like in RHIC or 
in LHC and find it too small to be detected. We further generalize the 
results for non-commutative Yang-Mills theories in diverse dimensions.
\end{abstract}

\newpage

\section{Introduction}

Jet quenching parameter \cite{Gyulassy:1993hr}, denoted by $\hat q$, 
is a measure of the radiative parton 
energy loss when they interact strongly with the medium (the quark-gluon 
plasma or QGP), produced in the ultrarelativistic heavy ion collision 
(for example, 
Au-Au collision in RHIC or Pb-Pb collision in LHC), as they traverse through
it (see, for example, \cite{Baier:2000mf} for reviews on the subject). 
$\hat q$ 
characterizes the properties of the medium and so by studying 
it one can gain insight about the properties of QCD matter in the extreme
conditions as produced in the collision process. The phenomenological models
of medium induced parton energy loss which account for the strong suppression 
of high-$p_T$ hadronic spectra observed in the experiment use a single 
jet quenching parameter, but this way it is model-dependent and it relies 
on the perturbative QCD framework \cite{Eskola:2004cr}. 

However, there are indications both from
experimental data \cite{Eskola:2004cr,Shuryak:2004cy} as well as
lattice QCD calculations \cite{Asakawa:2003re} that QGP, at
energies not far above the cross-over from the confinement (or 
the hadronic) phase,
is strongly coupled. A model-independent strong coupling calculation of $\hat
q$ has been performed in \cite{Liu:2006ug} using AdS/CFT correspondence 
\cite{Maldacena:1997re} and was found to
agree \cite{Liu:2006ug} reasonably well with the experimental result 
\cite{Eskola:2004cr} of RHIC. In this approach
one uses weakly coupled string or supergravity theory to calculate 
quantities in the strongly coupled gauge theory. In the gauge theory
$\hat q$ can be related to the expectation value of a particular light-like
Wilson loop. The latter quantity in string or supergravity theory can
be computed by extremizing the area of the string world-sheet whose boundary 
is the loop in question \cite{Maldacena:1998im}. Thus one obtains $\hat q$ 
in the strongly coupled
gauge theory in a model-independent way. In this spirit, $\hat q$ in many
different cases have been calculated in \cite{Herzog:2006gh}.

In this paper we compute the jet quenching parameter when
the plasma is described by hot non-commutative Yang-Mills theory using 
gauge/gravity duality. So, in this case the boundary theory is defined on
a manifold, where some of the space coordinates are non-commutative. The idea
that space-time\footnote{There is problem with the time coordinate being
non-commutative as it is believed to lead to a nonunitary theory 
\cite{Gomis:2000zz}.} 
coordinates can be non-commutative is not new and dates back to Heisenberg
and Pauli \cite{Heisenberg:1929xj},
who first proposed non-commutativity of space-time in order to remove
infinities in quantum field theory before renormalization was successful. 
However, it was
Snyder \cite{Snyder:1946qz} and then Connes \cite{Connes:1987rs} who took 
the idea seriously. Connes and Chamseddine \cite{Chamseddine:1991qh}
suggested non-commutative geometry as an alternative to Riemannian geometry
and obtained a variant of general relativity along with a standard model in
non-commutative space -- a true geometric unification. Although there are
various claims 
\cite{Carroll:2001ws,Anisimov:2001zc,Mocioiu:2000ip,Doplicher:1994tu} 
for the lower bound of the non-commutativity scale,
experimentally there is no evidence for or against it. In string 
\cite{Seiberg:1999vs,Maldacena:1999mh,Hashimoto:1999ut} or M-theory
\cite{Connes:1997cr}
non-commutative gauge theory arises quite naturally in a particular low energy
limit. Just like ordinary Yang-Mills theory appears on the boundary of
AdS-space, a low energy decoupling limit of D3-brane background,
a space-space non-commutative Yang-Mills theory appears in a particular low
energy limit of (D1, D3) bound state \cite{Breckenridge:1996tt} background of 
type IIB string theory. In this case the
world-volume of D3-brane contains a magnetic field which in the decoupling 
limit becomes asymptotically large \cite{Seiberg:1999vs} and makes some of 
the spatial directions non-commutative \cite{Chu:1998qz}. So, it would be
natural to ask how non-commutativity affects the jet quenching
parameter of the plasma described by ordinary Yang-Mills theory and this
is the question we would like to explore in this paper.   

Due to the star product and the associated complications, the perturbative
calculation of the jet quenching parameter in NCYM theory would be quite
involved. However, like in the commutative case, assuming the plasma to be
strongly coupled, the natural approach to do this calculation would be
the gauge/gravity duality once we know the gravity dual of NCYM theory.
We take this approach where the dual gravity background for NCYM theory is
known to be a particular decoupling limit of (D1, D3) bound state of type IIB
string theory. We use a string probe in this background and extremize the
Nambu-Goto string world-sheet action in a static gauge and by 
gauge/gravity duality obtain the expectation value of a particular 
light-like Wilson
loop \cite{Maldacena:1998im}. This, in turn, determines the jet quenching 
parameter \cite{Baier:2000mf} of the plasma of hot
NCYM theory. In obtaining the final form of $\hat q_{\rm NCYM}$ we need 
to regularize an
integral which we will discuss in appropriate place. We obtain $\hat 
q_{\rm NCYM}$ for
both small and large non-commutativity. For small non-commutativity, the jet
quenching parameter $\hat q_{\rm NCYM}$, on top of the value 
$\sim (\sqrt{\hat \lambda}
T^3)$, gets corrected by $\sim -(\hat \lambda^{3/2} T^7 \theta^2)$, where $T$ 
is the
temperature of the plasma, $\hat \lambda$ is the 't Hooft coupling in the
NCYM theory and $\theta$ is the non-commutativity parameter. So, the jet 
quenching (or the parton energy loss) gets reduced by the non-commutative 
effect
and also the reduction gets enhanced with increasing temperature as $T^7$. On
the other hand, for large non-commutativity the jet quenching varies as $\sim
1/(\sqrt{\hat \lambda} T \theta^2)$. So, the jet quenching in this case varies
inversely with temperature. We then generalize $\hat q_{\rm NCYM}$ for NCYM 
theories in other dimensions.

Jet quenching in (3+1)-dimensional NCYM theory has also been 
considered recently in \cite{Sadeghi:2010zp} and \cite{AliAkbari:2011fu}.
In \cite{Sadeghi:2010zp} a (3+1)-dimensional non-commutative, non-relativistic 
Yang-Mills theory has been considered following \cite{Panigrahi:2010cm}.
So, when the non-relativistic parameter is set to zero, the jet quenching
result must coincide with the result obtained in this paper. However,
the authors gave only a formal expression and did not discuss
the subtleties involved in the exact calculation performed in this paper.
On the other hand \cite{AliAkbari:2011fu} discussed NCYM theory from 
Sakai-Sugimoto
model and the background considered there is D4-brane instead of D3-brane.
However, the conclusion found there is quite similar 
to what we find here. 

This paper is organized as follows. In section 2, we compute the jet quenching
parameter in (3+1)-dimensional NCYM theory and the dual gravity background
is given by a decoupling limit of (D1, D3) brane bound state system. We
generalize this result to include the jet quenching calculation in other
dimensions in section 3. The dual gravity background in this case is given by
the NCYM decoupling limit of (D$(p-2)$, D$p$) brane bound state system.
Our conclusion is presented in section 4.  

\section{Jet quenching for (3+1)-dim. NCYM theory}

In this section we will calculate the jet quenching parameter for
(3+1)-dimensional NCYM theory from a light-like Wilson loop using
gauge/gravity duality. The dual gravity background for NCYM theory
is the NCYM decoupling limit of non-extremal (D1, D3) bound state system 
of type IIB string theory.  
The various field configurations of non-extremal (D1, D3) brane bound state 
solution are given as \cite{Cai:2000hn},
\bea\label{d1d3}
ds^2 &=& H^{-\frac{1}{2}}\left(-f(dt)^2 + (dx^1)^2 + \frac{H}{F}\left((dx^2)^2
+ (dx^3)^2\right)\right) + H^{\frac{1}{2}}\left(\frac{dr^2}{f} + r^2
d\Omega_5^2\right)\nn
e^{2\phi} &=& g_s^2 \frac{H}{F},\qquad B_{23} = \frac{\tan\alpha}{F}\nn
A_{01} &=& \frac{1}{g_s} (H^{-1}-1)\sin\alpha\coth\varphi,\qquad
A_{0123} = \frac{1}{g_s}\frac{(1-H)}{F}\cos\alpha \coth\varphi + {\rm T.\,T.}
\eea
where the various functions appearing above are,
\be\label{fns}
f = 1 - \frac{r_0^4}{r^4},\qquad H = 1 + \frac{r_0^4 \sinh^2 \varphi}{r^4},
\qquad F = 1 + \frac{r_0^4 \cos^2\alpha \sinh^2\varphi}{r^4}
\ee
The metric in \eqref{d1d3} is given in the string frame. Note that D3-branes
lie along $x^1,\,x^2,\,x^3$ and D1-branes lie along $x^1$. The angle $\alpha$
measures the relative numbers of D1 and D3-branes and is defined as,
$\cos\alpha = N/\sqrt{N^2+M^2}$, where $N$ is the number of D3-branes and $M$
is the number of D1-branes per unit co-dimension two surface transverse to
D1-branes. Also $\varphi$ is the boost parameter and $r_0$ is the radius of
the horizon of the non-extremal (D1, D3) bound state solution. $\phi$ is the
dilaton and $g_s$ is the string coupling constant. $B_{23}$ is the NSNS 2-form
which is responsible for the appearance of non-commutativity in the decoupling
limit. $A_{01}$ and $A_{0123}$ are the RR 2-form and 4-form respectively. 
T.T. denotes a
term involving the transverse part of the brane to make the corresponding
field-strength self-dual. However, we do not need its explicit form for our
discussion. 

The NCYM decoupling limit \cite{Seiberg:1999vs} is a low energy limit for 
which we zoom into the region
\be\label{ncym}
r_0 < r \sim r_0 \sqrt{\sinh \varphi \cos \alpha} \ll r_0 \sqrt{\sinh \varphi}
\ee
It is clear from \eqref{ncym} that $\varphi$ is very large and the angle
$\alpha$ is close to $\pi/2$. Now in this approximation \eqref{ncym}, we get
\be\label{approx}
H \approx \frac{r_0^4 \sinh^2\varphi}{r^4},\qquad \frac{H}{F} \approx
\frac{1}{\cos^2\alpha(1+a^4 r^4)} \equiv \frac{h}{\cos^2\alpha}
\ee
where,
\be\label{definition}
h =  \frac{1}{1+a^4 r^4}, \qquad {\rm with}, \quad a^4 = \frac{1}{r_0^4
  \sinh^2\varphi \cos^2\alpha}
\ee  
It is clear from \eqref{d1d3} that the asymptotic value of the $B$-field is
$\tan \alpha$ and since $\alpha \to \pi/2$ in the NCYM limit, the $B$-field
becomes very large. Note that the non-vanishing component of the $B$-field is
$B_{23}$ which gives rise to a magnetic field in the D3-brane world-volume and
is responsible for making $x^2$ and $x^3$ directions non-commutative.
Using \eqref{approx} we rewrite the metric in \eqref{d1d3}
in light-cone coordinate as,
\bea\label{lcmetric}
ds^2 &=& \frac{r^2}{r_0^2\sinh\varphi}\left[-(1+f)dx^+dx^- + \frac{1}{2}(1-f)
\left[(dx^+)^2 + (dx^-)^2\right] \right.\nn
& & \qquad\qquad + h \left[(dx^2)^2 +
  (dx^3)^2\right]\bigg] + \frac{r_0^2 \sinh\varphi}{r^2}\frac{dr^2}{f} 
+ r_0^2 \sinh\varphi d\Omega_5^2\nn
&=& G_{\mu\nu} dx^{\mu} dx^{\nu}
\eea 
where we have defined $x^{\pm} = (t \pm x^1)/\sqrt{2}$. The function $h$ is as
given in \eqref{definition} and also in writing \eqref{lcmetric} we have
rescaled the coordinates as, $x^{2,\,3} \to \cos\alpha\, x^{2,\,3}$. The above
metric, along with the other field configurations (in the NCYM limit) in
\eqref{d1d3}, is the gravity dual of (3+1)-dimensional NCYM theory at finite 
temperature. 

Now to compute the expectation value of the (light-like) Wilson loop (${\cal
C}$) we extremize the action ($S({\cal C})$) of the string world-sheet 
whose boundary is the mentioned loop \cite{Liu:2006ug,Liu:2006he}. 
The Nambu-Goto action for the string world-sheet is
\be\label{ngaction}          
S = \frac{1}{2\pi\alpha'}\int d\tau d\sigma \, \sqrt{{\rm det}\,g_{ab}}
\ee
where $g_{ab}$ is the induced metric on the world-sheet and is given as,
\be
g_{ab} = \frac{\partial x^{\mu}}{\partial \xi^a}\frac{\partial
  x^{\nu}}{\partial \xi^b} G_{\mu\nu}
\ee
Here $G_{\mu\nu}$ is the background metric given in \eqref{lcmetric} and
$\xi^{a,\,b}$, $a,\,b = 0,\,1$ are the world-sheet coordinates $\tau = \xi^0$
and $\sigma = \xi^1$. Due to reparametrization invariance of the action
\eqref{ngaction} we can set $\tau =
x^-$ and $\sigma = x^2$. The length of the rectangular loop ${\cal C}$ along 
$x^2$ and $x^-$ are $L$ and $L^-$ respectively and we assume $L^-\gg L$. As a
result the surface is invariant under $\tau$-translation and we have
$x^\mu (\tau,\sigma) = x^\mu (\sigma)$. Furthermore, the Wilson loop lies at
$x^+$ = constant and $x^3$ = constant. Note that one of the sides of the
rectangular Wilson loop is chosen along a non-commutative direction ($x^2$)
so that the jet quenching parameter evaluated from this Wilson loop will be 
affected by non-commutativity. The radial coordinate $r(\sigma)$ gives
the string embedding and we impose the condition that the world sheet 
has ${\cal C}$ as its
boundary, i.e., $r(\pm L/2) = r_0 \Lambda$, for some finite $\Lambda$. We will
take $\Lambda \to \infty$ at the end. The action \eqref{ngaction} now reduces
to, 
\be\label{action1}
S = \frac{\sqrt{2}L^-}{2\pi\alpha' \sinh\varphi} \int_0^{L/2} d\sigma 
\left[\frac{1}{1+a^4 r^4} +
  \frac{r_0^4\sinh^2\varphi}{r^4-r_0^4}(r')^2\right]^{\frac{1}{2}} 
\ee
where $r' = \partial_{\sigma} r$.
Defining new dimensionless variables $y=r/r_0$, $\tilde \sigma = 
\sigma/(r_0\sinh\varphi)$
and $\ell=L/(r_0\sinh\varphi)$, we can rewrite the action as,
\be\label{action2}
S = \frac{\sqrt{2} L^- r_0}{2\pi\alpha'}\int_0^{\ell/2} d\sigma 
\left[\frac{1}{1+a^4 r_0^4 y^4} + \frac{(y')^2}{y^4-1}\right]^{\frac{1}{2}}
\ee         
Note that in writing \eqref{action2} we have omitted the `tilde' from $\sigma$
for convenience. The equation of motion following from \eqref{action2} for
$y(\sigma)$ is given as,
\be\label{eos}
y' =
\left[1-q_0^2(1+a^4r_0^4y^4)\right]^{\frac{1}{2}}\frac{\sqrt{y^4-1}}{q_0(1+a^4
  r_0^4 y^4)}
\ee 
where $q_0$ is an integration constant. From the first factor in
\eqref{eos} we have $q_0 < 1/(1+a^4r_0^4y^4)^{\frac{1}{2}}$ for all values 
of $y$. In fact,
$q_0$ has more stringent restriction to be mentioned later. So, the above
equation has a solution\footnote{Note that here we discard another solution
at UV corresponding to the surface at infinity. Since $\hat q$ is a property 
of the thermal medium and does
not describe UV physics, the surface at infinity is not physically relevant
\cite{Liu:2006ug}.}
such that $y$ starts from $\Lambda$ and then comes
all the way down to $y=1$ where there is a turning point with $y'=0$ and goes
back again to $\Lambda$. Integrating \eqref{eos} we obtain,
\be\label{ell}
\ell = 2\int_0^{\ell/2} d\sigma = 2 q_0\int_1^\Lambda dy\,
\frac{1+a^4r_0^4y^4}{\sqrt{(y^4-1)
  \left[1-q_0^2(1+a^4r_0^4y^4)\right]}} 
\ee       
Now since $\ell = L/(r_0\sinh\varphi)$ is very small compared to any other
length scale of the problem, it implies from \eqref{ell} that $q_0$ must
be very small, i.e., $q_0 \ll 1/\sqrt{1+a^4r_0^4\Lambda^4}$ and so, we can 
expand 
\eqref{ell} in powers of $q_0$ and from there we formally obtain its value as,
\be\label{q0}
q_0 = \frac{\ell}{2}\left[\int_1^\Lambda dy \frac{1+a^4r_0^4y^4}{\sqrt{y^4-1}}
  \right]^{-1}
\ee   
Substituting \eqref{eos} in the action \eqref{action2} and then again
expanding in powers of $q_0$, we obtain
\be\label{action3}
S-S_0 = \frac{\sqrt{2}L^- r_0 q_0^2}{4\pi\alpha'}\int_1^\Lambda dy
\frac{1+a^4r_0^4y^4}{\sqrt{y^4-1}} = \frac{\sqrt{2}L^- r_0 \ell^2}
{16 \pi\alpha'}\left[\int_1^\Lambda 
dy \frac{1 + a^4r_0^4y^4}{\sqrt{y^4-1}}\right]^{-1}
\ee
where in the last expression we have used \eqref{q0}. In the above $S_0$
denotes the action for the world-sheet of two free strings (or the self
energy of the quark-antiquark pair). Here we would like to remark that
the integral in the square bracket in \eqref{action3} is actually divergent if
we take the boundary ($\Lambda$) where the NCYM theory lives to $\infty$.
The evaluation of the action here differs from
the commutative case. In the commutative case the action, after subtracting the
self-energy of the quarks, i.e., \eqref{action3} becomes finite and this can
be seen if we put $a^4 r_0^4$, which is a measure of non-commutativity (to be
discussed later), to zero. However, for the non-commutative case, the action
\eqref{action3} is still divergent if we put $\Lambda \to \infty$. The reason
for this divergence is that in the non-commutative case the NCYM theory does
not live at $ r=\infty$\footnote{This is implicit in the quark-antiquark
potential calculation done in \cite{Maldacena:1999mh} (see also 
\cite{Dhar:2000nj}). There it was not
possible to fix the position of the string at infinity since a small 
perturbation would change it violently. So, the calculation was performed by
going to a conjugate `momentum' variable and the energy was found to be
divergent. A finite answer was obtained only after subtracting the divergent
part. This, in turn, implies that the boundary screen is not at infinity but at
a finite radial distance \cite{Dhar:2000nj}.}, 
the usual boundary of the AdS$_5$-space, but
rather lives on a surface which is at a finite distance before
$r=\infty$. Instead of directly evaluating this 
distance what we will do is
that we will first evaluate the integral in the square bracket of
\eqref{action3} for finite $\Lambda$ and then subtract the part which is
divergent when
we put $\Lambda \to \infty$. This way we regularize the integral in order to
give any meaning to the extremized action\footnote{There are two ways to
describe the finiteness of the integral in the action \eqref{action3}. 
Either we take
$\Lambda$ to be finite in which case it is obviously finite (in this case
the integral can be evaluated only if we know the exact position of the
boundary)  or we take
$\Lambda$ to be infinite and subtract the unique divergent part (as explicitly
calculated below) of the integral and obtain a finite result. In the former
case the boundary is at a finite radial distance, but for the latter case it
is at infinity. But, effectively, they describe the same thing. Here we have
adopted the second approach.}. 
Once the subtraction is made
the NCYM theory can be considered to be living effectively on the usual 
$r=\infty$ boundary. So, we first evaluate the integral for finite $\Lambda$ 
as follows,
\bea\label{integral}
\int_1^{\Lambda} dy \frac{1+a^4r_0^4y^4}{\sqrt{y^4-1}} &=&
-\Lambda\sqrt{\Lambda^4-1} + \frac{1}{3}(3+a^4r_0^4) \frac{\sqrt{\pi} 
\Gamma\left(\frac{5}{4}\right)}{\Gamma\left(\frac{3}{4}\right)}\nn
& & +\frac{1}{3}(3+a^4r_0^4) \Lambda^3 \,
_2F_1\left(-\frac{3}{4},\frac{1}{2};\frac{1}{4};\frac{1}{\Lambda^4}\right)
\eea  
where $_2F_1(a,b;c;1/\Lambda^4)$ is a hypergeometric function. For large
$\Lambda$ it has an expansion
\be\label{expansion}
_2F_1\left(a,b;c;\frac{1}{\Lambda^4}\right) = 1 + \frac{ab}{c}
\frac{1}{\Lambda^4} + \frac{a(a+1)b(b+1)}{2c(c+1)}\frac{1}{\Lambda^8} + \cdots
\ee
Using the above expansion \eqref{expansion} in the rhs of \eqref{integral} and
finally setting $\Lambda \to \infty$, we find that apart from a finite part
the above integral has a single divergent piece of the form 
$(a^4r_0^4/3)\Lambda^3$ and all other terms vanish. So, removing the 
divergent part we get the regularized integral as,
\be\label{regintegral}
\int_1^\infty dy \frac{1+a^4r_0^4y^4}{\sqrt{y^4-1}} = \left(1+
  \frac{a^4r_0^4}{3}\right) \frac{\sqrt{\pi}\Gamma\left(\frac{5}{4}\right)}
{\Gamma\left(\frac{3}{4}\right)}
\ee
Substituting \eqref{regintegral} in \eqref{action3} yields,
\be\label{action4}
S-S_0 = \frac{\sqrt{2}L^- r_0 \ell^2}{16\pi\alpha'}
\frac{\Gamma\left(\frac{3}{4}\right)}{\sqrt{\pi}\Gamma\left(\frac{5}{4}\right)}
\left(1+ \frac{a^4r_0^4}{3}\right)^{-1}
\ee
Now to extract the jet quenching parameter $\hat q_{\rm NCYM}$ from 
\eqref{action4}, we use the definition,
\be\label{def}
e^{-2(S-S_0)} = \langle W({\cal C})\rangle = e^{-\frac{1}{4\sqrt{2}} \hat
  q_{\rm NCYM} L^- L^2}
\ee    
where $W({\cal C})$ denotes the Wilson loop and the factor 2 in front of
$(S-S_0)$ denotes that we are dealing with adjoint Wilson loop. Thus from
\eqref{action4} we have,
\be\label{jetq}
\hat q_{\rm NCYM} = \frac{r_0}{\pi \alpha' r_0^2 \sinh^2\varphi} 
\frac{\Gamma\left(\frac{3}{4}\right)}{\sqrt{\pi}\Gamma\left(\frac{5}{4}\right)}
\left(1+ \frac{a^4r_0^4}{3}\right)^{-1}
\ee
where we have used $\ell = L/(r_0\sinh\varphi)$. Note that here we have
expressed the jet quenching parameter in terms of the paramaters of string
theory or supergravity. However, since the jet quenching parameter is a 
property of a gauge theory, we must express it in terms of the parameters
of the NCYM theory and we can do that by using the gauge/gravity dictionary 
\cite{Maldacena:1999mh}. The temperature of the
non-extremal (D1, D3) bound state which by gauge/gravity duality is the
temperature of the NCYM theory can be obtained from the metric in \eqref{d1d3}
and has the form,
\be\label{temp}
T = \frac{1}{\pi r_0\cosh\varphi} \approx \frac{1}{\pi r_0\sinh\varphi}
\ee
where in the last expression we have used the fact that in the decoupling
limit \eqref{ncym}, $\varphi$ is large. Also from the charge of the D3-brane 
we have
\be\label{charge} 
r_0^4 \sinh^2\varphi = 2\hat \lambda \alpha'^2
\ee
Here $\hat \lambda = \hat g_{\rm YM}^2 N$ is the 't Hooft coupling of the 
NCYM theory and $\hat g_{\rm YM}$ is the NCYM coupling, with $N$ being the
number of D3-branes or the rank of the gauge group. The NCYM 't Hooft coupling
is related to the ordinary 't Hooft coupling by $\lambda = (\alpha'/\theta)
\hat \lambda$, where $\theta$ is the non-commutativity parameter defined by
$[x^2,\,x^3] = i \theta$. Here $\theta$ is a finite parameter and in the
decoupling limit as $\alpha' \to 0$, $\hat \lambda$ remains finite. 
Using \eqref{temp} and \eqref{charge} we obtain,
\be\label{tempcharge}
\sinh\varphi = \frac{1}{\pi^2 \sqrt{2\hat\lambda}\,T^2 \alpha'}, \qquad
{\rm and} \qquad r_0 = \pi  \sqrt{2\hat\lambda} \,T \alpha'
\ee
Also we have 
\be\label{ncparameter}
a^4 r_0^4 = \frac{1}{\sinh^2\varphi \cos^2\alpha} = \pi^4
(2\hat\lambda) T^4 \theta^2
\ee 
Note that in the above we have used the decoupling limit $\cos\alpha =
\alpha'/\theta$ and as $\alpha' \to 0$, $\alpha \to \pi/2$ as we mentioned
earlier. Also, from \eqref{ncparameter} we notice that since $a^2r_0^2$ is
proportional to $\theta$, therefore, $ar_0$ is a measure of non-commutativity.
Now using \eqref{tempcharge} and \eqref{ncparameter} in \eqref{jetq} we find
that for small non-commutativity ($ar_0 \sim \theta \ll 1$) the jet quenching 
parameter in the NCYM theory has the form,
\be\label{jetqsmall}
\hat q_{\rm NCYM} = \frac{\pi^{\frac{3}{2}} \Gamma\left(\frac{3}{4}\right)}
{\Gamma\left(\frac{5}{4}\right)}\sqrt{\hat\lambda} T^3\left[1 - \frac{\pi^4
     \hat\lambda T^4 \theta^2}{3} + O(\theta^4)\right]
\ee
We notice from \eqref{jetqsmall} that when the non-commutativity parameter 
$\theta$ is put to zero, we recover the jet quenching of the ordinary
Yang-Mills plasma (note that in this case the NCYM 't Hooft coupling $\hat
\lambda$ becomes equal to ordinary 't Hooft coupling $\lambda$ and also 
in writing \eqref{jetqsmall} we have replaced $2\hat\lambda$ by $\hat\lambda$
to match the commutative results in \cite{Liu:2006ug}.
This difference in a factor of 2 is just a convention as mentioned in
\cite{Liu:2006he}.). 
But in the presence of non-commutativity we find that the jet quenching gets 
reduced from its commutative value and the reduction gets enhanced with 
temperature as $T^7$, keeping the other parameters fixed. The reduction in
the jet quenching for the non-commutative case can be intuitively understood
as non-commutativity introduces a non-locality in space due to
space uncertainty and  there is no point-like interaction among the
partons. So, the parton energy loss would be less in this case.

We can try to estimate the correction (the second term in \eqref{jetqsmall})
in the jet quenching due to non-commutativity from the experimental bound on
the non-commutativity scale. In the literature various disparate experimental 
bounds on $\theta$ has been obtained from various physical considerations.
The bound on $\theta$ has been claimed to be $\sim$ 
(1 -- 10 TeV)$^{-2}$ in \cite{Carroll:2001ws}, whereas, 
it is $\sim (10^{12}$ -- $10^{13}$ 
GeV)$^{-2}$ in \cite{Anisimov:2001zc} or even stronger 
$\sim (10^{15}$ GeV)$^{-2}$ in \cite{Mocioiu:2000ip}. In theories of gravity 
it can be of the order of Planck scale $\sim 
(10^{19}$ GeV)$^{-2}$ \cite{Doplicher:1994tu}. It is clear that in all 
these cases except the first 
one there is no hope of getting a significant correction due to 
non-commutativity in collider experiments. At RHIC collision energy 
$\sim 200$ GeV where the temperature attained by QGP is $\sim 300$ MeV, 
even the first case does not give a significant correction ($\pi^4 \hat
\lambda T^4 \theta^2/3 \sim 4.96 \times 10^{-12}$ taking\footnote{We have
taken the 't Hooft coupling of the NCYM theory to be the same as that of the
commutative theory, although there is no concrete reason for this. This is
taken just for the estimate. Actually these two couplings are related as given
earlier and as $\alpha' \to 0$, $\lambda \to 0$, but $\hat \lambda$ remains 
finite. We have taken this finite value to be $6\pi$ for better comparison.}  
$\hat\lambda = 6\pi$
and $T = 300$ MeV relevant for the Au-Au collision at RHIC and taking
$\theta = 1$ TeV$^{-2}$) compared to the leading order term. At LHC where
the collision energy would be much higher, the temperature of the QGP may rise
and is expected to go up to (1 -- 10) GeV. In that case
the correction to
the jet quenching due to non-commutativity can be estimated to be
$\pi^4 \hat \lambda T^4 \theta^2/3 \sim$   
(6.12 $\times$ 10$^{-6}$ -- 6.12 $\times$ 10$^{-10}$), still too low to be
detected. Conversely, in order to get a 10$\%$ correction on the
jet quenching parameter due to non-commutativity, the temperature of the 
plasma would have to be $T \sim 200$ GeV.

For large non-commutativity ($ar_0 \sim \theta \gg 1$), on the other hand,
the jet quenching parameter \eqref{jetq} takes the form,
\be\label{jetqlarge}
\hat q_{\rm NCYM} = \frac{3 \Gamma\left(\frac{3}{4}\right)}
{\pi^{\frac{5}{2}} \Gamma\left(\frac{5}{4}\right)}\frac{1}{ \sqrt{\hat
    \lambda}\,T \theta^2}\left[1 - \frac{3}{\pi^4  \hat\lambda T^4
    \theta^2} + O\left(\frac{1}{\theta^4}\right)\right]
\ee
We thus find that for large non-commutativity, the jet quenching varies
inversely with temperature and also inversely with the square-root of the NCYM
't Hooft coupling. In this case we can not recover the commutative result as 
the non-commutativity parameter is large. This completes our derivation and
the discussion of the jet quenching parameter for the plasma of (3+1)-dimensional 
NCYM theory. 

\section{Jet quenching for NCYM theories in 
other dimensions}

We can easily generalize the above calculation of the jet quenching parameter 
in (3+1)-dimensions to NCYM theories in other dimensions. The procedure is
exactly similar to the (3+1)-dimensional case and so we will be brief here. For
generalization, the gravity background we use is (D$(p-2)$, D$p$) brane 
bound state system of type II string theory and is given as \cite{Cai:2000hn},
\bea\label{dpdp2}
ds^2 &=& H^{-\frac{1}{2}}\left(-f(dt)^2 + \sum_{i=1}^{p-2}(dx^i)^2 + 
\frac{H}{F}\left((dx^{p-1})^2
+ (dx^p)^2\right)\right)\nn
& & + H^{\frac{1}{2}}\left(\frac{dr^2}{f} + r^2
d\Omega_{8-p}^2\right)\nn
e^{2\phi} &=& g_s^2 \frac{H^{\frac{5-p}{2}}}{F},\qquad B_{p-1,p} = 
\frac{\tan\alpha}{F}\nn
A_{01\ldots p-2} &=& \frac{1}{g_s} (H^{-1}-1)\sin\alpha\coth\varphi,\qquad
A_{01\ldots p} = \frac{1}{g_s}\frac{(1-H)}{F}\cos\alpha \coth\varphi
\eea
where the various functions appearing above are,
\be\label{fns2}
f = 1 - \frac{r_0^{7-p}}{r^{7-p}},\qquad H = 1 + \frac{r_0^{7-p} 
\sinh^2 \varphi}{r^{7-p}},
\qquad F = 1 + \frac{r_0^{7-p} \cos^2\alpha \sinh^2\varphi}{r^{7-p}}
\ee
Here D$p$-branes lie along $x^1,\,x^2,\,\ldots,\,x^p$ and D$(p-2)$-branes lie
along $x^1,\,x^2,\,\ldots,\, x^{p-2}$. $\alpha$ and $\varphi$ are as defined
before, but now for D$(p-2)$, D$p$ branes. The NCYM decoupling limit is 
\cite{Alishahiha:1999ci,Roy:2009sw}
\be\label{ncymlimit}
r_0 < r \sim r_0 \sinh^{\frac{2}{7-p}}\varphi \cos^{\frac{2}{7-p}}\alpha 
\ll r_0
\sinh^{\frac{2}{7-p}} \varphi
\ee 
In this approximation,
\be\label{approx1}
H \approx \frac{r_0^{7-p} \sinh^2\varphi}{r^{7-p}}, \qquad
\frac{H}{F} \approx \frac{1}{\cos^2\alpha\left(1 + a^{7-p}{r^{7-p}}\right)}
\ee
where,
\be\label{approx2}
h = \frac{1}{1+a^{7-p} r^{7-p}}, \qquad {\rm with} \quad a^{7-p} =
\frac{1}{r_0^{7-p}\sinh^2\varphi \cos^2\alpha}
\ee
Now using \eqref{approx1} and \eqref{approx2} we can rewrite the metric
\eqref{dpdp2} in light-cone coordinates as follows,
\bea\label{lcmetric2}
ds^2 &=& \frac{r^{\frac{7-p}{2}}}{r_0^{\frac{7-p}{2}}\sinh\varphi}
\left[-(1+f)dx^+dx^- + \frac{1}{2}(1-f)
\left[(dx^+)^2 + (dx^-)^2\right] + \sum_{i=2}^{p-2}(dx^i)^2 \right.\nn
& & \qquad\qquad + h \left[(dx^{p-1})^2 +
  (dx^p)^2\right]\bigg] + \frac{r_0^{\frac{7-p}{2}} \sinh\varphi}
{r^{\frac{7-p}{2}}}\frac{dr^2}{f} 
+ \frac{r_0^{\frac{7-p}{2}}}{r^{\frac{3-p}{2}}} \sinh\varphi d\Omega_{8-p}^2\nn
&=& G_{\mu\nu} dx^{\mu} dx^{\nu}
\eea
In writing \eqref{lcmetric2} we have rescaled $x^{p-1,\,p} \to \cos\alpha
x^{p-1,\,p}$ and these are the two noncommutative directions with 
$[x^{p-1},\,x^p] = i \theta$. This metric along with the other field
configurations in the NCYM limit is the gravity dual of $(p+1)$-dimensional 
NCYM theory. It should be noted here that $p$ in the above configuration
\eqref{lcmetric2} is greater than 2 and since the brane configuration in
general develops instability for $p>4$, so, strictly speaking, the results 
below are sound only for $p=3,4$.  
Now setting $\tau = x^-$ and $\sigma = x^p$, we can evaluate the 
Nambu-Goto action exactly as before and we obtain in this case,
\be\label{action5}
S-S_0 = \frac{\sqrt{2}L^- r_0 q_0^2}{4\pi\alpha'}\int_1^\Lambda dy
\frac{1+a^{7-p}r_0^{7-p} y^{7-p}}{\sqrt{y^{7-p}-1}} = 
\frac{\sqrt{2}L^- r_0 \ell^2}
{16 \pi\alpha'}\left[\int_1^\Lambda
dy \frac{1 + a^{7-p}r_0^{7-p}y^{7-p}}{\sqrt{y^{7-p}-1}}\right]^{-1}
\ee
Again the integral in the square bracket is divergent for $\Lambda \to \infty$.
So, as before, we regularize it and the regularized integral has the value,
\be\label{regintegralp}
\int_1^{\infty} dy \frac{1+a^{7-p} r_0^{7-p} y^{7-p}}{\sqrt{y^{7-p} -1}}
= \left(1 + \frac{2 a^{7-p} r_0^{7-p}}{9-p}\right)a_p, \quad {\rm with}
\quad a_p = \frac{\sqrt{\pi}}{7-p}\frac{\Gamma\left(\frac{5-p}{2(7-p)}\right)}
{\Gamma\left(\frac{6-p}{7-p}\right)}
\ee
Using \eqref{regintegralp} in \eqref{action5} and also using the definition 
\eqref{def} we extract the form of the jet quenching parameter for the plasma of
$(p+1)$-dimensional NCYM theories as,
\be\label{jetqp}
\hat q_{\rm NCYM} = \frac{r_0}{\pi \alpha' r_0^2 \sinh^2\varphi\,a_p}
\left(1+\frac{2 a^{7-p}r_0^{7-p}}{9-p}\right)^{-1}
\ee 
Now to express $\hat q_{\rm NCYM}$ in terms of the parameters of the NCYM
theory we use the following relations 
\cite{Roy:2009sw,Itzhaki:1998dd,Alishahiha:1999ci},
\be\label{tempchargep}
T = \frac{7-p}{4\pi r_0\sinh\varphi}, \quad 
r_0^{7-p}\sinh^2\varphi = d_p \hat \lambda \alpha'^{5-p}, \quad {\rm with}
\quad d_p = 2^{7-2p} \pi^{\frac{9-3p}{2}}\Gamma\left(\frac{7-p}{2}\right)
\ee
Using \eqref{tempchargep} in \eqref{jetqp} we find the form of the jet
quenching parameter as,
\be\label{jetqp1}
\hat q_{\rm NCYM} = \frac{8\sqrt{\pi} \Gamma\left(\frac{6-p}{7-p}\right)}
{\Gamma\left(\frac{5-p}{14-2p}\right)} b_p^{\frac{1}{2}} T^2 \left(
  \sqrt{\hat \lambda}\, T\right)^{\frac{2}{5-p}}\left[1 + \frac{8\pi^2}{9-p}
b_p T^2 \left(\sqrt{\hat \lambda}\,T\right)^{\frac{4}{5-p}}\theta^2\right]^{-1}
\ee
where $b_p = \left[\left\{2^{16-3p} \pi^{(13-3p)/2} 
\Gamma((7-p)/2)\right\}/(7-p)^{7-p}\right]^{2/(5-p)}$. 
From the above expression it may seem that since the jet quenching 
parameter in $(p+1)$-dimensions depends, in general, on $p$, so it is
non-universal. However, by defining a $T$-dependent dimensionless effective 
't Hooft
coupling $\hat \lambda_{\rm eff}(T) = \hat \lambda T^{p-3}$ \cite{Liu:2006he}
we can rewrite the
jet quenching parameter for small non-commutativity as,
\bea\label{jetqp2}
\hat q_{\rm NCYM} &=& \frac{8\sqrt{\pi} \Gamma\left(\frac{6-p}{7-p}\right)}
{\Gamma\left(\frac{5-p}{14-2p}\right)} b_p^{\frac{1}{2}} 
\hat\lambda_{\rm eff}^{\frac{p-3}{2(5-p)}}(T) 
\sqrt{\hat \lambda_{\rm eff}(T)}\,T^3\nn
& &\times \left[1 - \frac{8\pi^2}{9-p} b_p \hat\lambda_{\rm eff}^{\frac
{p-3}{5-p}}(T)
\hat\lambda_{\rm eff}(T) T^4 \theta^2 + O(\theta^4)\right]\nn
&=& \frac{8\sqrt{\pi} \Gamma\left(\frac{6-p}{7-p}\right)}
{\Gamma\left(\frac{5-p}{14-2p}\right)} \sqrt{\hat a(\hat \lambda_{\rm eff})}
\sqrt{\hat \lambda_{\rm eff}(T)}\,T^3
\left[1 - \frac{8\pi^2}{9-p} \hat a(\hat \lambda_{\rm eff})
\hat\lambda_{\rm eff}(T) T^4 \theta^2 + O(\theta^4)\right]\nn
\eea
where in writing the second expression above we have defined a new parameter
$\hat a(\hat\lambda_{\rm eff})\equiv b_p \hat\lambda_{\rm eff}^{(p-3)/(5-p)}$.
The quantity $\hat a$, as has been mentioned for the ordinary Yang-Mills 
theory, 
characterizes the number of degrees of freedom at temperature $T$ since both
the energy density and the entropy density can be seen to be proportional 
to it \cite{Liu:2006he}. 
Thus by comparing the second expression of \eqref{jetqp2}
with \eqref{jetqsmall}, we find that the dependence of the jet quenching 
parameter on the coupling constant, the temperature and the non-commutativity
parameter is indeed universal. This has been noted for $(p+1)$-dimensional 
ordinary Yang-Mills theory in \cite{Liu:2006he} and we see here that it 
holds for $(p+1)$-dimensional NCYM theory as well. 
From \eqref{jetqp2}, we notice that when $\theta=0$, we recover precisely the
commutative result obtained in \cite{Liu:2006he,Chakraborty:2011ah}. 
Non-commutativity reduces the value of 
the jet quenching parameter from its commutative value. On the other hand for
large non-commutativity we get,
\bea\label{jetqp3}
\hat q_{\rm NCYM} &=& \frac{(9-p) \Gamma\left(\frac{6-p}{7-p}\right)}
{\pi^{\frac{3}{2}}\Gamma\left(\frac{5-p}{14-2p}\right)} \frac{1}
{b_p^{\frac{1}{2}}
\hat\lambda_{\rm eff}^{\frac{p-3}{2(5-p)}}(T)
\sqrt{\hat \lambda_{\rm eff}(T)}\,T \theta^2}\nn
& & \times
\left[1 - \frac{9-p}{8\pi^2} \frac{1}{b_p \hat\lambda_{\rm eff}^{\frac{p-3}
{5-p}}(T) \hat\lambda_{\rm eff}(T) T^4 \theta^2} + O\left(\frac{1}
{\theta^4}\right)\right]\nn
&=& \frac{(9-p) \Gamma\left(\frac{6-p}{7-p}\right)}
{\pi^{\frac{3}{2}}\Gamma\left(\frac{5-p}{14-2p}\right)} \frac{1}
{\sqrt{\hat a(\hat \lambda_{\rm eff})}
\sqrt{\hat \lambda_{\rm eff}(T)}\,T \theta^2}\nn
& &\times\left[1 - \frac{9-p}{8\pi^2} \frac{1}{\hat a(\hat \lambda_{\rm eff})
\hat\lambda_{\rm eff}(T) T^4 \theta^2} + O\left(\frac{1}
{\theta^4}\right)\right]
\eea
Again comparing the second expression of \eqref{jetqp3}
with \eqref{jetqlarge}, we find that the jet quenching parameter has a 
universal structure for all $p$.

\section{Conclusion}

To conclude, in this paper we have calculated the jet quenching parameter of
the plasma of (3+1)-dimensional NCYM theory using gauge/gravity duality. We
used the fundamental string as a probe in the background of (D1, D3) bound
state system in the NCYM decoupling limit and extremized the Nambu-Goto 
action in a particular light-cone gauge. This by gauge/gravity duality is
related to a light-like Wilson loop of the NCYM theory where the boundary 
of the string world-sheet is the loop in question. From the Wilson loop we 
extracted the jet quenching parameter. However, to obtain its value we had to
regularize an integral. This is on top of another regularization that is
usually performed in the commutative case corresponding to the self-energy
of the quark-antiquark pair. In the non-commutative case another 
regularization is necessary because here the NCYM theory does not live at the
usual boundary $r=\infty$, but at a finite distance before $r=\infty$
\cite{Maldacena:1999mh}. Once we
remove the divergent term, NCYM theory can be considered to be living on
$r=\infty$ boundary. We gave its expression for both small and large
non-commutativity. For small non-commutativity the jet quenching parameter
gets reduced 
in value from its commutative counterpart as non-commutativity introduces a
non-locality in space and smoothes out the interaction. The correction term 
is proportional to
$T^7$ and so the reduction gets enhanced with increasing temperature as $T^7$.
We have also estimated the correction in the jet quenching parameter due to
non-commutativity by taking into account the experimental bound on the
non-commutativity scale $\theta$. We found that it would be difficult to find
any significant contribution due to non-commutativity at the present collider
energy both at RHIC or at LHC. So, the contribution due to 
non-commutativity will be significant only at much higher energies. 
We have generalized our results for plasma of NCYM theories in diverse
dimensions. We found that the jet quenching parameter typically
depends on $p$, however, it has a universal structure if we rewrite it in
terms of a quantity $\hat a(\hat \lambda_{\rm eff})$, which characterizes the
number of degrees of freedom at temperature $T$.        
  
\vspace{.2cm}

\section*{Acknowledgements}

One of us (SR) would like to thank 
Kamal L Panigrahi for discussion and collaboration at an early stage of 
this work. We would also like to thank Munshi G Mustafa for useful discussions.
We wish to thank the anonymous referee for comments which helped us, 
we hope, to improve the manuscript.

\end{document}